\documentclass[traditabstract]{aa}
\usepackage{graphicx}
 \usepackage{lscape}
 \usepackage{colortbl}
 \usepackage{natbib}
 \usepackage{verbatim} 
  \usepackage{comment} 
\usepackage{ulem}
\usepackage{url}
 \bibpunct{(}{)}{;}{a}{}{,}
\usepackage{txfonts}
\begin{document}
\input psfig.tex
\def\simgt{\stackrel{>}{{}_\sim}}
\def\simlt{\stackrel{<}{{}_\sim}}

\titlerunning{Richness-mass relation self-calibration for galaxy clusters
}
\title{Richness-mass relation self-calibration for galaxy clusters
}
\author{S. Andreon
\inst{1},
\and J. Berg\'e
\inst{2}}{
\institute{
$^1$ INAF--Osservatorio Astronomico di Brera, Milano, Italy\\
$^2$ ETH Zurich, Department of Physics, Wolfgang Pauli Strasse 27, 
8093 Zurich, Switzerland \\
\email{stefano.andreon@brera.inaf.it}
}
\date{Received date; accepted date}
\abstract{
This work attains a threefold objective: first, we derived the richness-mass scaling in
the local Universe from data of 53 clusters with individual measurements of mass. 
We found a $0.46\pm0.12$ slope and a $0.25\pm0.03$ dex
scatter measuring richness with a previously developed method. Second, we showed
on a real sample of 250 $0.06<z<0.9$ clusters, most of which are at $z<0.3$, 
with spectroscopic redshift
that the colour of the red sequence allows
us to measure the clusters' redshift to better than $\Delta z=0.02$. Third,
we computed the predicted prior of the richness-mass scaling to forecast the
capabilities of future wide-field-area surveys of galaxy clusters
to constrain cosmological parameters. To this aim, 
we generated a simulated universe obeying the richness-mass
scaling that we found. We observed it with a PanStarrs 1+Euclid-like survey, allowing for
intrinsic scatter between mass and richness, for errors on mass, on richness, 
and for photometric redshift errors. We fitted
the observations with an evolving five-parameter richness-mass scaling 
with parameters
to be determined. Input parameters were recovered, but only if the
cluster mass function and the
weak-lensing redshift-dependent selection function were accounted 
for in the fitting of the mass-richness scaling. This emphasizes the
limitations of often adopted simplifying assumptions, such as 
having a mass-complete redshift-independent sample. 
We derived the uncertainty and the covariance matrix of the
(evolving) richness-mass scaling, which are the input ingredients
of cosmological forecasts using cluster counts. 
We find that the richness-mass scaling parameters can
be determined $10^5$ better than estimated in previous works that did 
not use weak-lensing mass estimates, 
although we emphasize that this high factor was derived with
scaling relations with different parameterizations. 
The better knowledge of the scaling parameters likely has  
a strong impact on the relative importance of
the different probes used to constrain cosmological parameters.
The fitting code used for computing the predicted prior, 
inclusing the treatment of the mass function and of the weak-lensing
selection function, is provided in the appendix. It can be
re-used, for example, to derive the predicted prior of other
observable-mass scalings, such as the $L_X$-mass relation.
}
\maketitle

\section{Introduction}

If one has a sample of $N$ clusters with measured
properties, $obsn_i$, $z_i$ (where $i =1,2,...,N$), 
for example in a Euclid-like survey, their constraints 
on the cosmological parameters
$\theta=(\Omega_M,\Omega_\Lambda,\sigma_8,w,...)$
can be derived by 
applying Bayes's theorem to obtain the posterior
distribution of the cosmological parameters,
\begin{equation}
p(\theta|obsn_i,z_i)\propto p(obsn_i,z_i|\theta) p(\theta) \quad ,
\end{equation}
where $p(\theta)$ is the prior on cosmological parameters (e.g.
from other surveys) and $p(obsn_i,z_i|\theta)$ is the likelihood
of measuring $N$ clusters with measured properties $obsn_i$, $z_i$. 
If the mass $M$ were observable, the cosmological
parameters $\theta$ would be constrained by fitting $p(M,z|\theta)$ 
to the observed distribution. 
However, this direct fit is not possible with survey data, 
because one needs to rely on an observable (mass proxy), 
such as richness or $Y_X$,  and fit the distribution of
the observable, $O$ with $p(O,z|M)$. To
estimate the cosmological parameters, one needs to assume 
a model for the scaling between the mass and the observable (usually a power law) and some
knowledge about how precisely the parameters describing this
relation are known (the mass-observable prior): the knowledge may
range from very precise (a delta function prior) to
very uncertain (e.g. an improper uniform prior). 
Of course, cosmological estimates benefit from
better known scaling parameters, i.e. priors that enclose
a narrow volume of the parameter space that describes the mass-observable 
scaling.

Most previous forecasts dealing with counts of galaxy clusters 
(e.g. Lima \& Hu 2005, Sartoris et al. 2010, Carbone et al. 2012)
assumed the precision with which the parameters of the mass-observable
scaling will be known instead of measuring it. One of the
purposes of this work is to quantify this part of the inference step:
we aim to compute the uncertainties of the mass-observable scaling,
i.e. the volume of the mass-richness scaling parameter space enclosed
by the posterior probability distribution. We consider, specifically,
cluster richness as the mass proxy. This analysis gives us the input prior of 
cosmological forecasts using cluster counts. 

The paper is organized as follow:
in Sect. 2 we measure the mass-observable
relation in the local Universe from real data, we determine
how well the cluster redshift can be inferred from the colour of the
red sequence and we compute in
which part of the universe the observable can be measured with
current data. In Sect. 3
we assume a fiducial model where the relation between mass and proxy
does not evolve. 
We populate an (simulated) observable universe, we measure the
parameter uncertainties by fitting an evolving mass-observable relation
to all data (real and simulated), and we test our ability
to recover an evolving mass-observable relation.
Finally, in Sect.~4 we discuss our results and compare the
measured uncertainties of the mass-observable scaling with what has
been thus far assumed in cosmological forecasts. Sect. 5 summarizes
the results of this work.

Throughout this paper we assume $\Omega_M=0.3$, $\Omega_\Lambda=0.7$, 
$H_0=70$ km s$^{-1}$ Mpc$^{-1}$, $\sigma_8=0.8$. 
Magnitudes are quoted in their
native system (quasi-AB for SDSS magnitudes).  All logarithms in
this work are on base ten, unless otherwise indicated. 
All quantities are measured at the $r_{200c}$ radius, whose enclosed
averaged mass density is 200 times the critical density. The
richness-mass calibration in this paper refers to richnesses measured following 
the Andreon \& Hurn (2010)
prescriptions, and therefore cannot be used
for other types of richnesses, e.g. Abell (1958) richnesses.
We adopt the standard statistical
notation: the $\sim$ symbol reads ``is drawn from" or ``is distributed as"
and  the $\leftarrow$ symbol reads ``take the value of".

\section{Calibration of the mass-proxy from current data}

\begin{figure}
\psfig{figure=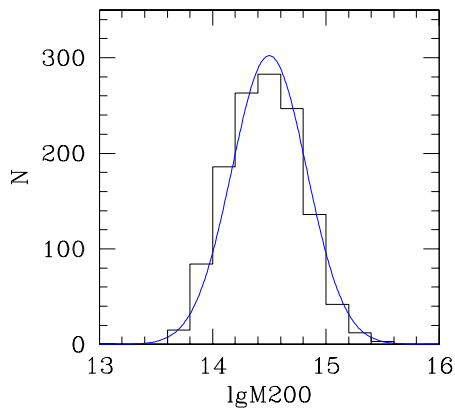,width=6truecm,clip=}
\caption[h]{Computed selection function (histogram) and
its adopted Gaussian approximation (curve).
}
\label{fig:fig2}
\end{figure}

\begin{figure}
\psfig{figure=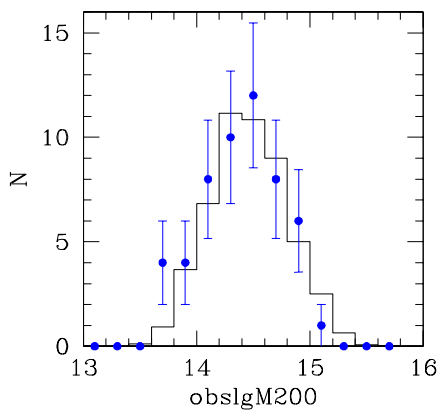,width=6truecm,clip=}
\caption[h]{Distribution of the expected $obslgM200$ fake data 
(histogram) and distribution of real data (points). 
Errorbars mark count standard deviation (i.e. are $\sqrt{n}$), 
not the error.
}
\label{fig:fig2}
\end{figure}

\subsection{Local calibration of the richness-mass relation based on real data}

In this section,
we are interested in the scaling between richness and mass in the
local Universe taking into account the noise in their measurement and
selection effects. 

We re-analysed the very same data that were used in 
Andreon \& Hurn (2010), adopting the modeling appropriate
for the task of current interest. In short, the data consist of
cluster richnesses, $n200$, based on red galaxies measured on
specified luminosity and colour ranges within a fiducial radius,
and masses derived from the caustic technique 
computed using 208 galaxies on average per cluster for
53 galaxy clusters at $0.03 < z < 0.1$. As detailed in Andreon \& Hurn 
(2010), the parameters describing the mass-richness relation
do not change if we use instead velocity-dispersion-based masses. 
We emphasize that we used the values denoted with a hat in
Andreon \& Hurn (2010) because they are
derived without knowledge of the mass-related quantities
($r_{200}$), precisely like in real survey data.
For notation simplicity, we here suppress the hat
notation adopted there. 

Because it is an X-ray selected sample, the considered cluster sample is 
controlled, not random; therefore, bright clusters are over-represented.
In general, a non-random selection causes
biases in the recovered regression parameters if the 
selection is neglected (Gelman et al. 2003; Stanek et al. 2006; 
Pacaud et al. 2007; Andreon, Trinchieri \& Pizzolato 2011;
Andreon \& Moretti 2011; Andreon \& Hurn 2012; 
and see also sec 3.2  where
we discuss this problem at length for a sample for which the non-random 
selection cannot be ignored).
To be precise, the studied cluster sample is a random
sampling (as detailed in Andreon \& Hurn 2010) of an X-ray selected
sample. Its controlled nature allows us to
compute the mass selection function, which is essential,
in general, to correct for non-random mass selection leading to
biases in the recovered regression parameters. 
We computed the mass selection function (mass prior)} as follows:  we
assumed that the local cluster
mass function is described by a Jenkins et al. (2001) mass function
at the masses of interest ($\log M> 13.5$ $M_\odot$). Our results
are independent of the chosen parametrization  (e.g. 
if Press \& Schecther (1974)
would be adopted). We then followed 
Stanek et al. (2006): the mean relation between the
X-ray luminosity and the mass has a slope
equal to $1.59$, intercept equal to $lnLx_{15}=1.34$ 
(in a system employing different Hubble constant conventions for
luminosity and mass), intrinsic scatter of $0.59$, 
and the distribution of the (neperian ln) X-ray
luminosity at a given mass is Gaussian, i.e.
\begin{eqnarray}
lnL_{X,i} &\sim& \mathcal{N}(1.59 \ (lgM200_i-15)  +1.34, 0.59^2) \quad .
\label{eqn:eqn88}
\end{eqnarray}

This\footnote{The tilde symbol indicates  
a similarity subject to stochasticity, either because of noise or because
of intrinsic differences among members. Broadly speaking, the 
tilde symbol 
indicates that we account for uncertainty or non-homogeneity (variety).}
allows us to populate a simulated local universe, $0.03<z<0.1$,  with
clusters of X-ray luminosity $lnL_{X,i}$. The flux of
these (simulated) clusters is computed and the objects are 
kept in the sample if $f_X>3 \ 10^{-12}$ erg s$^{-1}$ cm$^{-2}$, which is the
flux threshold adopted by Rines \& Diaferio (2006), the parent sample from which Andreon
\& Hurn (2010) studied a random subsample. Fig. 1 shows the result of this simulation, and
the adopted analytic (Gaussian) parametrization:
\begin{eqnarray}
lgM200 &\sim& \mathcal{N}(14.5,0.33^2) \quad .
\label{eqn:eqn8}
\end{eqnarray}

Assuming Eq. 3, we computed the expected
distribution of the observed values of lgM200, $obslgM200$ of our
simulated survey, assuming a common error for the mass error, 
$0.14$ dex, the average value of the studied sample. We compared
this to the actual observed distribution (i.e. real data) 
in Fig 2. The agreement is impressive (there are no free parameters
to tune),
showing that our modelling of the selection function captures 
the data behaviour and gives us $p(lgM200)$ i.e. the probability that
a cluster has mass $lgM200$ and is included in the sample (i.e. the mass
prior). The derived $p(lgM200)$ allows us to avoid the biases 
coming from the non-random mass distribution of our sample.

We proceed by specifying the assumed mathematical dependence between 
the quantities involved in our problem.
We need to acknowledge the 
uncertainty in all measurements and therefore,
because of errors, observed and true values
are not identically equal.
The variables $n200_i$ and $nbkg_i$ represent the true richness 
and the true background galaxy counts in the studied solid angles.
We measured the number of galaxies in both cluster and control field
regions, $obstot_i$ and $obsbkg_i$ respectively, for each of our 53
clusters (i.e. for $i=1,\ldots,53$).
We allowed Poisson errors for both and we assumed that
all measurements are conditionally independent.
The ratio between the cluster and control field solid angles,
$C_i$, is known exactly. In formulae:
\begin{eqnarray}
obsbkg_i &\sim& \mathcal{P}(nbkg_i) \\
obstot_i &\sim& \mathcal{P}(nbkg_i/C_i+n200_i) \quad .
\end{eqnarray}

For each cluster, we have a cluster 
mass measurement and a measurement of the
error associated with this mass, $obslgM200_i$ and $obserrlgM200_i$
respectively.
We allowed Gaussian errors on mass:
\begin{equation}
obslgM200_i \sim \mathcal{N}(lgM200_i,obserrlgM200_i^2) \quad .
\label{eqn:eq9} 
\end{equation}

We assume a power law relation between mass and $n200$ 
with intercept $\alpha+1.5$, slope $\beta$ and
intrinsic scatter $\sigma_{scat}$:
\begin{equation}
lgn200_i \sim
\mathcal{N}(\alpha+1.5+\beta (\log(M200_i)-14.5), \sigma_{scat}^2)
\label{eqn:eqn12}
\end{equation}

The quantity $\log(M200)$ is centred at an average value of 14.5 and
$\alpha$ is centred at 1.5, for computational advantages in
the MCMC algorithm used to fit the model (it speeds up
convergence, improves chain mixing, etc.) and to reduce the covariance
between parameters. The relation is between true values, not
between observed values, which may be biased.

The priors on the slope and the intercept of the regression line in
Equation 7 were taken to be quite flat,
a zero mean Gaussian with very large variance for $\alpha$ and a
Students-$t$ distribution with one degree of freedom for $\beta$.
The latter choice was made to avoid that properties of galaxy clusters 
depend on astronomer rules of measuring angles (from the $x$ or from the $y$ axis).
This agrees with the model choices in
Andreon (2006 and later works). Our $t$ distribution on $\beta$ is 
mathematically equivalent to a uniform prior on the angle $b$. In formulae:
\begin{eqnarray}
\alpha &\sim& \mathcal{N}(0.0,10^4) \\
\beta &\sim& t_1 \quad .
\label{eqn:eqn111}
\end{eqnarray}

For the true values of the background,
we choose to impose no strong a-priori values, only enforcing
positivity, by adopting an improper uniform prior,
\begin{eqnarray}
nbkg_i &\sim& \mathcal{U}(0,\infty) \quad .
\label{eqn:eqn8}
\end{eqnarray}

\begin{figure}
\centerline{\psfig{figure=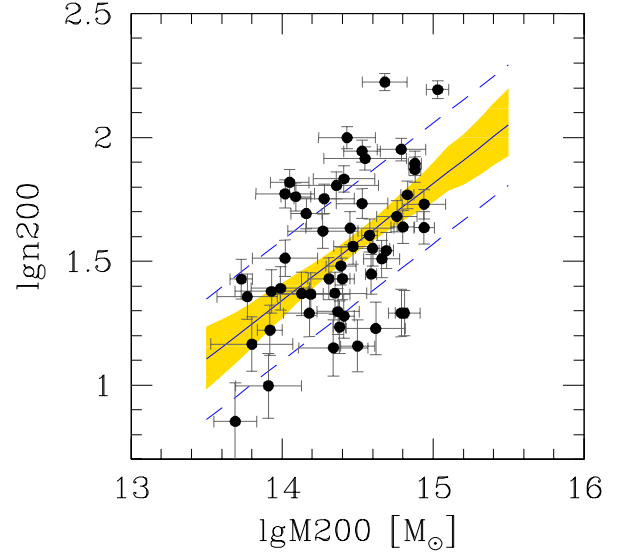,width=8truecm,clip=}}
\caption[h]{Richness-mass scaling.
The solid line marks the mean fitted regression line of $log(n200)$ on $lgM200$, while the dashed lines
show  this mean plus or minus the intrinsic scatter $\sigma_{scat}$. The shaded region marks the 68\%
highest posterior 68 \% credible interval for the regression. Error bars on the data points represent observed
errors for both variables. The distances between the data and the regression line is due in part to the
measurement error and in part to the intrinsic scatter.}
\label{fig:fig1}
\end{figure}

\begin{figure*}
\centerline{\psfig{figure=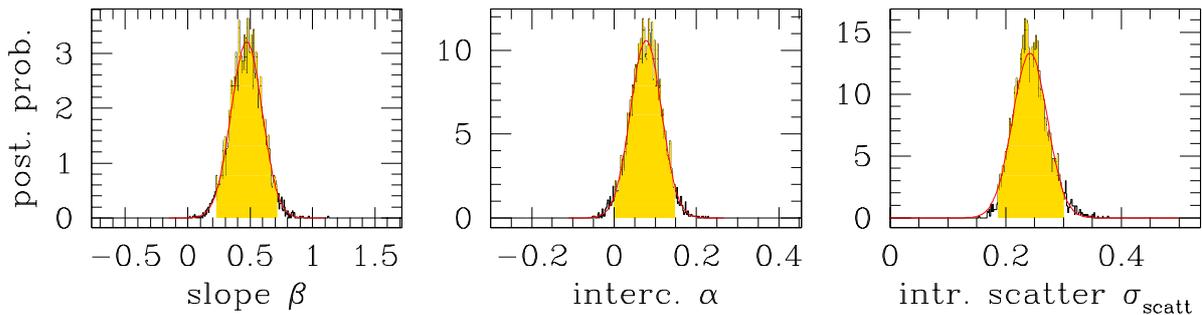,width=16truecm,clip=}}
\caption[h]{Posterior probability distribution for the
parameters of the richness-mass scaling computed from the real data.
The black jagged histogram shows the posterior as computed
by MCMC, marginalised over the other parameters. The red curve
is a Gaussian
approximation of it.  The shaded (yellow) range shows
the 95\% highest posterior credible interval. 
}
\end{figure*}

Fitting our sample of 53 clusters with the model above, we found:
 
\begin{equation}
lgn200 = (0.47\pm0.12) \ (lgM200 -14.5) +1.58\pm0.04  \quad .
\end{equation}
Unless otherwise stated, the results of the statistical computations 
are quoted in the form $x\pm y$ where $x$
is the posterior mean and $y$ is the posterior standard deviation.
All statistical computations were performed using JAGS (Plummer 
2010), see the appendix for an example.

Figure 3 
shows the scaling between richness and mass, the observed
data, the mean scaling (solid line), and its 68\% uncertainty (shaded yellow
region) and the mean intrinsic scatter (dashed lines) around
the mean relation.  The $\pm 1$ intrinsic scatter band 
contains 60 \% of the data points and is 
not expected to contain 68\% of them, because of 
measurement errors.

Figure 4 
shows the posterior marginals for
the key parameters, i.e. for the intercept, slope, and intrinsic
scatter $\sigma_{scat}$. These marginals
are reasonably well approximated by Gaussians.
The intrinsic mass scatter at a given richness, 
$\sigma_{scat}=\sigma_{lgM200|\log n200}$, is small, $0.25\pm0.03$ dex.
These posterior probability distributions are
dominated by the data (their widths are much smaller than the prior
widths), i.e. our results are independent of the assumed prior
to all practical effects.
Parameters show no appreciable covariance (figure not shown) 
because of our choice
of zero-pointing masses near the data average (eq 7). This allows
a simpler summary of the posterior, which we 
use in our next inference step (eq 17 to 19).

We note that these results
are almost indistinguishable from results we might obtain without
modelling the selection function, basically because the prior 
is broad compared to $lgM200$ errors.

\subsubsection{Side comments}

Cosmological forecasts dealing with cluster
counts in the optical sometimes use the scatter
between observable and mass from Rykoff et al. (2012)  or
Rozo et al. (2009).
It is worth emphasizing that to measure the scatter
between two quantities, it is strongly preferable to have both.
Neither of these two works
have individual values of cluster masses.

It is worth remembering that the slope of the direct relation is not 
the inverse of the slope
of the inverse relation, i.e. if $O
\propto M^\gamma$, then usually $M \not \propto O^{1/\gamma}$ 
(e.g. Isobe et al. 1990, Andreon \& Hurn 2010).
Therefore, it is not surprising that the slope between mass and
richness is not the reciprocal of the slope 
determined in Andreon \& Hurn (2010)
for the inverse relation using the very same data. Furthermore, the slope
depicted in Figure 3 is not ``too shallow" compared to the data, a steeper slope
would systematically over- or under-estimate the cluster richness (see Andreon \& Hurn 
2010, 2012 for a brief astronomical introduction on regression fitting).

\subsection{In which part of the Universe is richness measurable with current data?}

The cluster richness was derived using
$g-r$ colour and luminosities
of galaxies brighter than an evolving limiting magnitude $M^e_V<-20$.

Figure 5 illustrates how depth and colour constraints change with redshift.
The top panel illustrates the apparent luminosity of a red $M^e_V=-20$ mag
galaxy, modelled as a $z_f=5$ single stellar population using the 2007
version of the Bruzual \& Charlot (2003) synthesis population model for
different filters: $g$, $r$, $i$, and $z$ for the $3\pi$ Steradian 
PanStarrs 1 survey 
(PS1, hereafter) and $riz$, $Y$, $J$, and $H$ for Euclid\footnote{http://www.euclid-ec.org} (Laureijs et al. 2011) with the
corresponding $\sim10 \sigma$ depth (horizontal ticks). For the $3\pi$ 
PanStarrs 1, 
we took the current depth, i.e. that already achieved after the first two
years of observation (Kaiser, N., private communication). The PS1 has a Y-like
filter, not plotted because it is shallower than the Euclid Y.  The
Dark Energy Survey
(Abbott et al. 2005) is deeper than PS1, but covers a smaller solid angle.
The Euclid consortium plan to have ground based $griz$ data deeper than 
our need over the whole 15000 deg$^2$ survey area (Laureijs et al. 2011).

\begin{figure}
\centerline{%
\psfig{figure=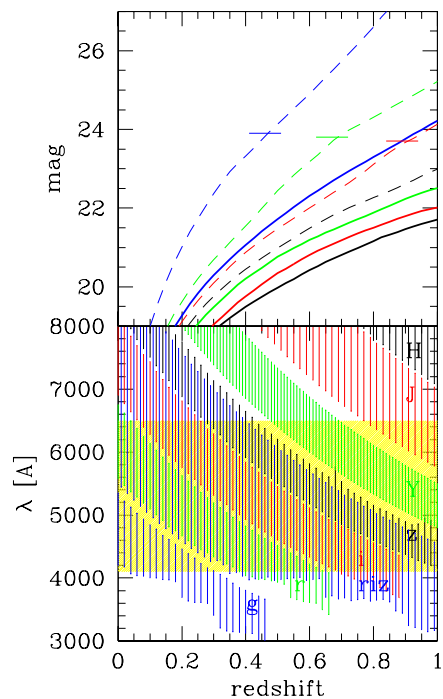,width=6truecm,clip=}
}
\caption[h]{Depth and wavelength coverage of the two-year PS1 and Euclid surveys. 
{\it Upper panel:}
$g$, $r$, $i$, and $z$ (from left to right) filters are indicated with dashed (blue, green,
red, and black) lines. 
$riz$, $y$, $J$, and $H$ (from left to right) Euclid filters are indicated with 
thick solid (blue, green,
red, and black) lines. The horizontal tick indicates the $\sim 10 \sigma$ depth, most
of them are at $z>1$ and thus not visible in the plot.
{\it Bottom panel:} Wavelength coverage of the filters for redshift bins where galaxies are brighter 
than the $\sim10 \sigma$ depth. The shaded (yellow) region marks the wavelength sampling
of $g-r$ at $z\sim 0$.}
\label{fig:fig3}
\end{figure}

The bottom
panel illustrates the wavelength range sampled by these filters. Only 
redshift bins where galaxies are brighter than the 10 $\sigma$ depth are plotted. 
The shaded yellow is the $\lambda$ range sampled by $g-r$ at 
$z<0.08$. As the figure shows, we always have at least two filters
in the shaded region, i.e. up to $z=1$ at least
these data have appropriate depth and wavelength coverage to count galaxies.
Indeed, the  $M^e_V<-20$ mag cut was chosen to precisely perform
this measurement on ten-year old MOSAIC-II CTIO images up to $z=0.82$ (e.g. 
those in Andreon et al.
2004a). These depths are routinely achieved in current surveys, such as the 
CFHTLS (Cuillandre \& Bertin 2006). 

To summarize, incoming (and also current)
surveys have the depth and filter coverage adequate to compute
the number of red galaxies needed to derive $n200$. Furthermore,
Andreon (2012) showed
that the galaxy background ($nbkg_i/C_i$ in eq. 5) is negligible even at
magnitudes fainter than those adopted in this work, and not detrimental
at all for the derivation of the cluster richness.

\begin{figure}
\psfig{figure=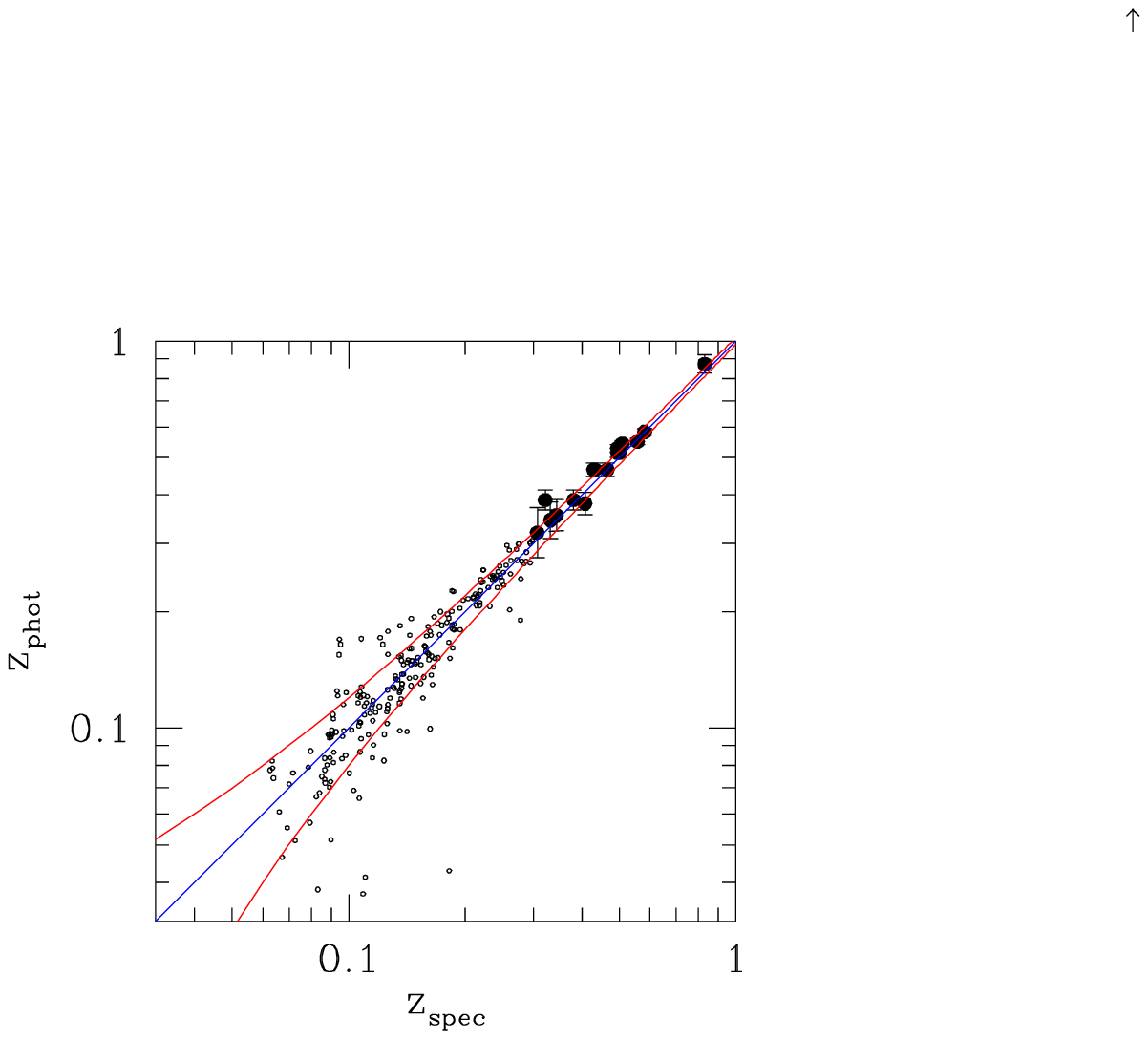,width=8truecm,clip=}
\caption[h]{Red sequence photometric redshift performance.
Spectroscopic redshift vs photometric redshift from the colour
of the red sequence in $g-r$ (small open points,
error bars are not plotted to avoid crowding, there are 228 plotted
points, sometime one on top of the other) or $R-z'$
(solid points with error bars).
The $z_{phot}=z_{spec}$ line and the $z_{phot}=z_{spec}\pm0.02$ loci
are indicated with solid lines.}
\label{fig:figz}
\end{figure}

\subsection{Which precision for photometric redshift?}

Surveys such as those performed by PanStarrs 1, DES, or 
Euclid will detect
thousands of clusters and it is unreasonable to expect that all of 
them will
have a spectroscopic redshift. How precise will their redshift estimate be?
We can set a conservative estimate
by considering current shallower surveys that sample similar redshifts.

We considered spectroscopic and
photometric redshifts of
the sample of 228 clusters at $0.06<z<0.3$ in 
Andreon (2003a,b) and the 16 $0.3<z<0.9$ clusters in Andreon et al. 
(2004a, 2004b).
They are all colour-detected with the red-sequence method of Andreon 
(2003a),
which is an adaptation of the Gladders \& Yee (2000) original method
(see Andreon 2003a for details) in either the SDSS early-data
release area or in the XMM-LSS field. These
clusters tend to be of
low richness and therefore to have a less prominent red sequence 
than that of the massive clusters that we consider below.
For both samples, the colour of the red sequence was determined
using two-band photometry only, $g-r$ (at $z<0.3$) or $R-z'$ (at $z>0.3$). The
photometric redshift was derived from the colour of
the red sequence adopting a relation between
redshift and colour (an empirical template at $z<0.3$, an old
galaxy template at higher redshift, as detailed in Andreon 2003a and Andreon et 
al. 2004a, respectively).
Fig 6 shows $z_{phot}$ vs $z_{spec}$ for the 244 clusters,
the (straight line) $z_{phot}=z_{spec}$ line and the $z_{phot}=z_{spec}\pm0.02$ loci.
Twenty-five percent of the points have $|z_{phot}-z_{spec}|>0.02$, while
$>32$ \% are expected if the photometric error is $0.02$. 
Even restricting the attention to $z>0.3$, 6 clusters
show $|z_{phot}-z_{spec}|>\sqrt{0.02^2+err^2_{z_{phot}}}$ vs 5.1
expected cases if the redshift derived from the red sequence
has an intrinsic scatter of $0.02$. This implies
that we can already achieve a $\Delta z =0.02$ precision using 
the colour of the
red sequence using two bands. Similar results were
found by Puddu et al. (2001) for a small, but X-ray selected (and therefore
more massive) cluster sample, and by High et al. (2010) for a small,
but mostly at $z>0.3$, cluster sample. In both cases the estimate of
the clusters' redshift is based on the colour of the red sequence.

The extremely good performance of the red sequence colour as a redshift
indicator is hardly surprising because of the implicit selection of one
single type of galaxies with a distinctive 4000 \AA \ break 
(spectrophotometric
bright early-type galaxies) and of the colour homogeneity of the
early-type galaxy class (e.g. Stanford, Eisenhardt, \& Dickinson 1998,
Kodama et al. 1998, Andreon 2003a,b, Andreon et al. 2004a).

In summary, we can safely assume for future clusters a (conservative) 
$0.02$ error on cluster (photometric) redshifts, 
because this performance is already achieved today using the 
colour of the red sequence. 

\section{Calibration with future surveys}

\subsection{Generation of mock-calibration Euclid data}

We generated a Monte-Carlo
simulated universe obeying to the mass-richness
scaling we just computed and observed it with a PanStarrs 1+Euclid-like survey. Our fiducial
universe has unevolving parameters that describe the  mass-richness scaling.  A
Euclid-like survey is needed to measure cluster masses, whereas for the computation of
cluster richness one needs shallower, but multicolour data, such as 
already acquired by the PanStarrs 1 survey. 

We followed Berg\'e et al. (2010) to compute the number (the probability
times the volume) of clusters in the Euclid-wide survey at redshift $z$,
with mass $lgM200$, which produces a weak-lensing signal with a given
signal-to-noise ratio $S/N$. We used the halo model with an NFW  (Navarro,
Frenk \& White 1997) profile, a Jenkins (2001) mass function, and a modified
Sheth, Mo \& Tormen (2001) bias (see the Berg\'e et al. 2010 appendix for
a detailed description). We assumed a galaxy shape
noise $\sigma_{\rm int} = 0.3$, and a galaxy number density $n_g = 30$
arcmin$^{-2}$. We also assumed that all halos are spherical and 
therefore did not account for the shape bias described by Hamana et al.
(2012). Projection effects are, in these observational
conditions and for clusters as massive as those of interest in our paper,
largely sub-dominant (Marian \& Bernstein 2006), and were neglected for this reason.
For the Euclid survey, we adopted the updated sky coverage (15000
deg$^2$). The
iso-density contours in Fig. 6 indicate lines where we expect 1,
10, and 100 clusters with an $S/N>5$ per bin of 0.1 dex in mass and
0.0275 in redshift in the Euclid survey (15000 square degrees). 
The minimal $S/N=5$
mass, lgM200trunc, is well described by
\begin{equation}
lgM200truc = 13.9891+1.04936 z+0.488881 z^2 \quad .
\end{equation}
We exploited these
masses to calibrate the richness-mass relation and its evolution.

First of all, we generated a Monte-Carlo 
realization of the  Berg\'e et al. (2010)
distributions. Then, we selected $S/N>5$ detections only, because we did not want
to deal with too noisy measurements (Hamana et al. 2012; Pace et al. 2007). 
Furthermore, we removed clusters at $z<0.03$ to avoid very nearby
clusters with galaxies bright and large, whose photometry will likely be 
corrupted\footnote{For example in the SDSS, which is much shallower and
therefore less tailored for faint galaxies, photometry of galaxies at $z<0.02$
suffers from shredding problems.}. This left us about 11000 clusters with
available $z_i,lgM200_i$, and $(S/N)_i$.

Cluster masses were then observed, i.e.
mass errors were taken to be Gaussian and equal to $errlgM200_i=\frac{1}{S/N}/ln(10)$, 
where the latter term is due to our choice of measuring errors
using decimal logarithms:
\begin{eqnarray}
obslgM200_i &\sim& \mathcal{N}(lgM200_i,errlgM200^2_i) \quad .
\label{eqn:eqn8}
\end{eqnarray}

Cluster richnesses were assigned to simulated clusters assuming the 
model measured in the local Universe, i.e. (Sec 2.1)
\begin{eqnarray}
lgn200_i &\sim& \mathcal{N}(0.47\ (lgM200_i-14.5)+1.58,0.25^2) \quad .
\label{eqn:eqn8}
\end{eqnarray}

We emphasize, once more, that we allowed for an intrinsic scatter, i.e. we
allowed clusters of a given mass 
to have a variety of
richnesses. Richnesses
were then observed: richness, as all measurements in this paper, have
errors, which were assumed to be Poissonian, 
\begin{eqnarray}
obslgn200_i &\sim& \mathcal{P}(lgn200_i) \quad .
\label{eqn:eqn8}
\end{eqnarray}

Finally, we also allowed Gaussian photometric errors, taken to be 0.02 at 
all redshifts (see Sec 2.3): 
\begin{eqnarray}
obsz_i &\sim& \mathcal{N}(z,0.02^2) \quad .
\label{eqn:eqn8}
\end{eqnarray}

This procedure yielded 10714 clusters with measured
$obslgn200_i,obslgM200_i,errlgM200_i$, and $obsz_i$, 
that we used to determine the relation between richness and mass and
its evolution. Fig 7 depicts them individually (points). Because
of mass errors, there are points below the minimal $S/N=5$ mass (the
diagonal slightly bent line).  The average richness ($obsn200$) 
of the simulated sample is 46 galaxies, 
while the median is 38 galaxies. Two thirds of them are at $z<0.3$,
where the scatter between redshift and photometric redshift is better
sampled by our real data (sec 2.3). 
The generated sample does not contain any cluster with
a weak-lensing detection at $S/N>5$ at $z>0.62$ (Fig 7).

\begin{figure}
\centerline{\psfig{figure=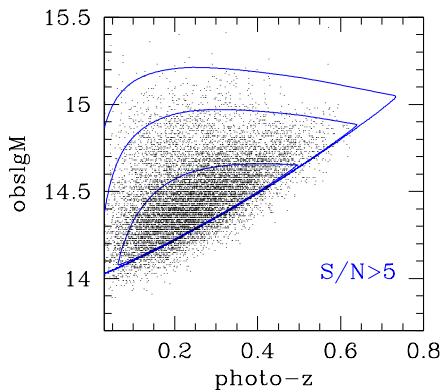,width=6truecm,clip=}}
\caption[h]{Contours: number of clusters for which weak-lensing mass estimates
can be obtained by a Euclid-like survey. From outer to inner
contours the lines represent isocontours of $S/N>5$ weak-lensing
detection of $1,10$, and $100$ clusters as a function of 
redshift and mass. Points: a Poisson realization of the above, with
errors on mass and redshift (these move points outside the $N=1$
contour).
}
\label{fig:fig5}
\end{figure}

\subsection{Determining the richness-mass predicted priors}

We now combined the real data from the local Universe with the simulated
data (depicted in Fig. 7), 
to compute how well we are able to measure
the richness-mass scaling at all redshifts. 
In this section we
will not use true values because these are unknown
for the real data. Furthermore, we cannot assume to know how
the parameters of the richness-mass scaling evolve, because
this is precisely what we want to infer from the data.

The information encoded in the local Universe (sect 2.1)
is the current prior:
\begin{eqnarray}
\sigma_{intrscat} &\sim& \mathcal{N}(0.25,0.03^2) \\
\alpha &\sim& \mathcal{N}(0.08,0.04^2) \\
\beta &\sim& \mathcal{N}(0.47,0.12^2) \quad .
\label{eqn:eqn8}
\end{eqnarray}

We assumed that the scatter 
and the intercept may both change with redshift:
\begin{eqnarray}
lgn200m_i &\leftarrow& \alpha+1.5 +\beta (lgM200_i-14.5) + \gamma \ln(1+z_i) \\
lgn200_i &\sim& \mathcal{N} (lgn200m_i, \sigma^2_{intrscat}(z_i)) \\
\sigma^2_{intrscat}(z_i) &\leftarrow& \sigma^2_{intrscat}-1 + (1+z_i)^{2\zeta}  \quad .
\label{eqn:eqn8}
\end{eqnarray}

While the adopted modelling of the evolution is common in previous
works (e.g. Sartoris et al. 2010, Carbone et al. 2012), we emphasize that a different 
modelling is possible and legitimate. We also emphasize that, as in
previous works, we assumed to perfectly known the analytic expression 
of the distribution function of the intrinsic scatter term (a Gaussian), 
when its shape should be left more flexible, or at very least, 
checked with data, because this uncertainty may be dominant
(Shaw et al. 2010). Equation 21 and the fitting code (given in
the appendix) may be easily modified replacing the
adopted Gaussian with a more flexible
distribution, e.g. by a mixture of two Gaussians, which guarantee a valid
(positive) probability distribution, unlike the Edgeworth series expansion
proposed in Shaw et al. (2010).

We adopted weak priors for the newly introduced parameters:
as prior for the $\gamma$ and $\zeta$ slopes we adopted a 
Students $t$ distribution centred on zero with one degree of freedom,
as for the slope $\beta$ in sec 2.1, to make our choice
independent of astronomer rules of measuring angles. In
formulae:
\begin{eqnarray}
\gamma &\sim& t_1 \\
\zeta&\sim& t_1 \quad .
\label{eqn:eqn8}
\end{eqnarray}

As in previous sections, richness has Poisson errors:
\begin{eqnarray}
obsn200_i &\sim& \mathcal{P} (n200_i) \quad ,
\end{eqnarray}
whereas masses and photometric redshifts have Gaussian errors:
\begin{eqnarray}
obslgM200_i &\sim& \mathcal{N}(lgM200_i,errlgM200^2_i) \\
obsz_i &\sim& \mathcal{N}(z,0.02^2) \quad .
\end{eqnarray}

To complete the model description, we need to specify the
mass prior. We cannot ignore that the mass function is
steep and that the weak-lensing $S/N>5$ cut introduces an abrupt discontinuity:
ignoring them would lead to a biased fit (the recovered
slope would be much shallower than the input one) due to 
a Malmquist-like bias. Indeed, mass errors tend to make the distribution
in mass broader, especially at low-mass values, because of
the sharp $S/N=5$ weak-lensing detection requirement, 
but also at high-mass values because of the steepness
of the mass function. Since high-mass values are overestimated and
low-mass values are underestimated, any quantity that is fitted
against these (biased) values neglecting the selection function
would return a shallower relation (see also Andreon \& Hurn 2010
for the similarly biased mass-richness relation of Johnston et al. 2007).
For mathematical simplicity and given the small mass range explored, 
we modelled the Jenkins et al. (2001) mass distribution at a given
redshift as a Schechter (1976) function with slope $-1$
and characteristic mass given by
\begin{equation}
lgM200^* = 12.6- (z-0.3) 
\end{equation}
truncated at $lgM200truc$, given by eq. 12. The parameters of eq 28 were
determined by fitting the Jenkins et al. (2001) mass function.

On the other hand, we do not need to model the optical cluster selection 
function, because the large cluster richness and the photometric 
depth allow all clusters that produce a detectable weak-lensing signal
to be easily detectable as overdensities of
red galaxy because they have, on average, 38 galaxies projected on a background of
(nearly) zero galaxies.

We do not need, either, to accurately model the redshift prior, because
photometric redshifts are well-determined. We can therefore assume an uniform
distribution for it
\begin{eqnarray}
z_i &\sim& \mathcal{U}(0,1) \quad .
\end{eqnarray}
although we emphasize that for large photometric redshift errors 
one should account for gradients in $n(z)$.

\begin{figure*}
\psfig{figure=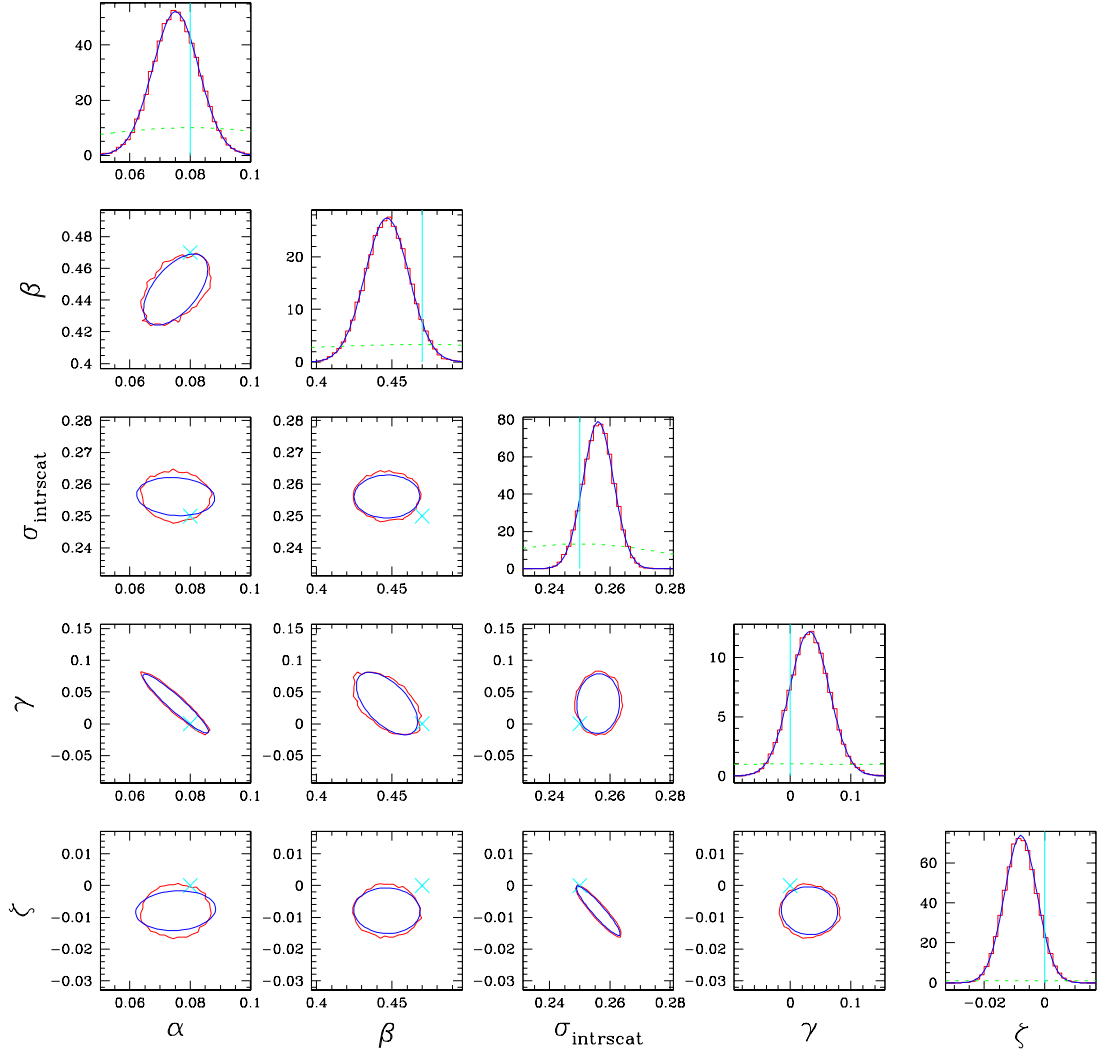,width=16truecm,clip=}
\caption[h]{Marginal (panels on the diagonal) and joint (other panels)
probability distributions of the mass-richness
scaling derived from real and simulated data for a PS1+Euclid-like
survey. Red jagged contours and histograms refer to probabilities
computed from the MCMC
sampling, while the blue smooth contours/lines refer to a Gaussian
approximation of the numerically found posterior. Contours are at 68\% 
probability level. Vertical (cyan) lines and crosses indicate the values used to
generate the data, while the dashed (green) lines show the current 
low-redshift calibration of the richness-mass scaling. 
}
\end{figure*}

We emphasize that modelling the mass- and selection function is compulsory; not
accounting for it would lead to a  fitted slope $\gg 5 \sigma$ different from
the input one. Therefore, results based on methods that do not allow for the
accounting of the mass- and selection function, e.g. the usual linear regression analysis
based on BCES (Akritas \& Bershady 1996), or simplistic forecast analyses
lacking any treatment of the selection function (as is typical of Fisher
analyses), should be used with great caution. On the other hand,
one should not be overly anxious about modelling the 
mass- and selection function:
what matters is their general shape, which drives the
correction of the bias, 
not their precise shape, i.e.  whether the mass function is a Tinker 
et al. (2008) or Jenkins et al. (2001) mass function, for instance. 
The uncertainty on the precise
shape of the mass function, neglected in this work because of the 
small considered mass range,
is an uncertainty of secondary importance compared to the large uncertainy
involved through the mass errors.
The main point to keep in mind is that the mass function is certainly not uniform,
it is evolving with redshift, the clusters
entering in the sample are not a random sampling of the mass function
(all those with low mass are excluded, and the limiting mass is changing
with redshift) and we
account for that (not accounting leads to parameters off by $\gg 5 \sigma$,
as mentioned),
while other observable-mass fitting models (sometime implicitely) assume a
uniform prior on cluster mass and mass-random selection, 
unless differently specified.

The software implementation of this fitting model is given in the appendix.

Fitting the simulated+real data with this model returns parameters
whose (posterior) probability distributions are depicted in Fig. 8. 
Fig. 8 and its summary in Table 1 are one of the main results of
this work, since they are the priors (starting points) needed to
forecast cosmological parameters with cluster data.

Marginal probabilities are shown on the diagonal, while the other
panels show the joint probability distributions, i.e. the covariance
between pairs of parameters. Each panel reports two closely packed 
lines: the red one is the Laplace (Gaussian) approximation of the
posterior, while the histogram/jagged contour is the straight
outcome of the numerical computation (somewhat noisy because of the
finite length of the MCMC chain). The Laplace (Gaussian)
approximation captures the probability distributions well.

The diagonal panels also show the input values (vertical lines). 
They are all within $1.5$ posterior standard deviations from the
recovered value\footnote{There is
only a 10 \% probability that in a five parameter fit 
all fitted values are found within 1 $\sigma$ from the input values,
and a 50 \% probability that they are all within 1.5 $\sigma$.}. 
By fitting the observed data, we recover the five parameters that
describe the mass-richness scaling with good accuracy
and without bias.

In addition to input
values, the diagonal panels show the current low-redshift
calibration of the richness-mass scaling (dashed green line). 
Euclid  masses significantly improve the current 
low-redshift calibration of the richness-mass scaling:
the intercept $\alpha$, currently known to within 10 \% ($0.04$ dex, sec
2.1), will be known with  a per cent  accuracy, 
the slope $\beta$,
currently known with a sizeable uncertainty ($0.47\pm0.12$, sec 2.1) will
have its uncertainty reduced by a factor $10$. The intrinsic scatter,
currently known with a $\sim10$ \% accuracy (sec 2.1), will be known with
a per cent accuracy. The evolution of the intrinsic scatter  and of the
intercept will be known with a $0.03$ and $0.005$ uncertainty,
respectively.  The computed posterior is 
$\sim10^3$ times narrower (in the $\alpha-\beta-\sigma$ space) 
than the
current calibration of the richness-mass scaling, a
significant improvement over the current low-redshift calibration. 
This capability makes the Euclid mission unique
and independent of the success of  observations other than the already
acquired PanStarrs 1 multicolour data. Instead, the  calibration of the
mass-proxy relation of the XXLS cluster survey (Pierre et al. 2011) must
rely on the success of an expensive XMM calibration program (Pierre et al.
2011), which is not yet implemented.   Similarly, the SPT survey
requires an external calibration. Although the current clusters sample consists 
only of 100 clusters (Reichardt et al. 2012), the currently available
calibration, not the sample size, is the main source of uncertainty 
in cosmological estimates.

There is a strong covariance between the evolution and the
$z=0$ value of the intercept ($\gamma-\alpha$ panel of Fig 8). 
It can be easily understood by noting that
$z=0$ is outside the range of sampled redshifts. The covariance between
intrinsic scatter and its evolution ($\zeta - \sigma_{intrscat}$ panel of Fig 8)
has a similar origin: the intrinsic
scatter is defined at an un-observed redshift, $z=0$, instead of a
redshift where it is well observed. 

Fig 9 compares the model fit (solid line) to the true input relation 
in stacks of 201 clusters  per point. The model fit on noisy data
and the (unobserved and unused
in the analysis) noise-less data agree well, indicating that the fit to the noisy
data captures the real trend of the noise-less true data well.

In summary, by fitting observed data we recover with good accuracy
and without bias the five parameters describing the mass-richness scaling.
In
particular, we assumed no evolution (i.e. $\gamma=0$ and $\zeta=0$) and 
recovered it, even allowing evolution on both scatter and intercept.
We will be able to measure the mass-richness scaling 
with an error (posterior parameter standard deviation) of $0.007$, $0.014$, 
$0.005$, $0.033$, and $0.005$ in $\alpha$, $\beta$, intrinsic scatter, $\gamma$, and
$\zeta$, respectively. These are the predicted prior widths of cosmological
forecasts. Table 1 lists the covariance matrix.

Strictly speaking, conclusions of this sub-section only hold if
our modelling of the richness-mass scaling is a reasonable approximation
of the scaling  
in the real Universe. Therefore it does not hold if,
for example,  the richness-mass scaling suddenly disappears
in the real Universe at $z=0.3$, for instance.

\begin{figure} \psfig{figure=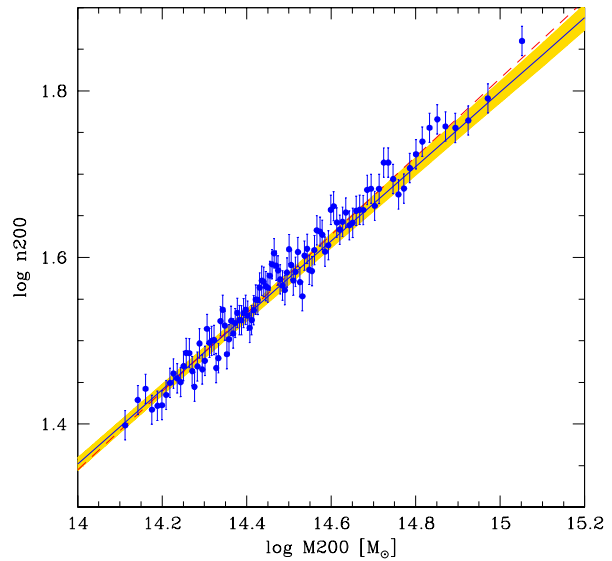,width=8truecm,clip=} 
\caption[h]{Richness-mass
scaling for the simulated PS1+Euclid-like data. The solid line marks the 
regression line fitted on observed data. 
The shaded region marks the 68\% highest posterior
credible interval for the regression. 
The red dashed line indicates
the input relation. The data points are
stacks of true data in bins of 201 clusters each, 
true data were never used in the fitting.} 
\label{fig:fig1}
\end{figure}

\subsection{What happens if $\sigma_{intrscat}$ doubles by $z=0.6$?}

To understand how the predicted prior is sensitive to a possible evolving
mass-proxy scaling, we generated new data with $\zeta=0.18$, i.e. generated from
a relation whose intrinsic scatter is twice as large at $z=0.6$ as at $z=0$.
To this aim, we replaced equation 14 by

\begin{eqnarray}
lgn200_i &\sim& \mathcal{N}(0.47\ (lgM200_i-14.5)+1.58,\sigma^2_{intrscat}(z_i)^2) \\
\sigma^2_{intrscat}(z_i) &\leftarrow& 0.25^2-1 + (1+z_i)^{2\zeta} \\
\zeta &\leftarrow& 0.18
\label{eqn:eqn88}
\end{eqnarray}

and re-generated the new (simulated) data. We fitted real+simulated data with
no change whatsoever, and, as for a non-evolving intrinsic
scatter, we recovered the input parameters, finding $\zeta=0.16\pm0.01$ (vs input
$\zeta=0.18$). The other four parameters were all recovered to better than
their uncertainty. More precisely, we found an error of $0.01$, $0.02$, 
$0.008$, $0.046$, and $0.010$ in $\alpha$, $\beta$, intrinsic scatter, $\gamma$,
and $\zeta$, respectively. These are larger (1.5 times, on average) than
before because with the larger scatter (at high redshift) more data are
needed to measure the mean relation with the same precision. Nevertheless,
the parameter volume they encompass is only a factor nine larger than
for a non-evolving intrinsic scatter, a negligible factor (a mere $0.01$ per
cent) compared to what we discuss below.  Marginal and joint
probability distributions (i.e. the covariance matrix and Fig 8 revised for these data)
are qualitatively similar, apart from the obvious $\approx 1.5$ factor.

In summary, an increasing intrinsic scatter, if present, would be easily recovered
from the data, with only a mild degradation of the overall performances
(a factor 9 for a five dimensional volume), and no bias.

\begin{figure*}
\psfig{figure=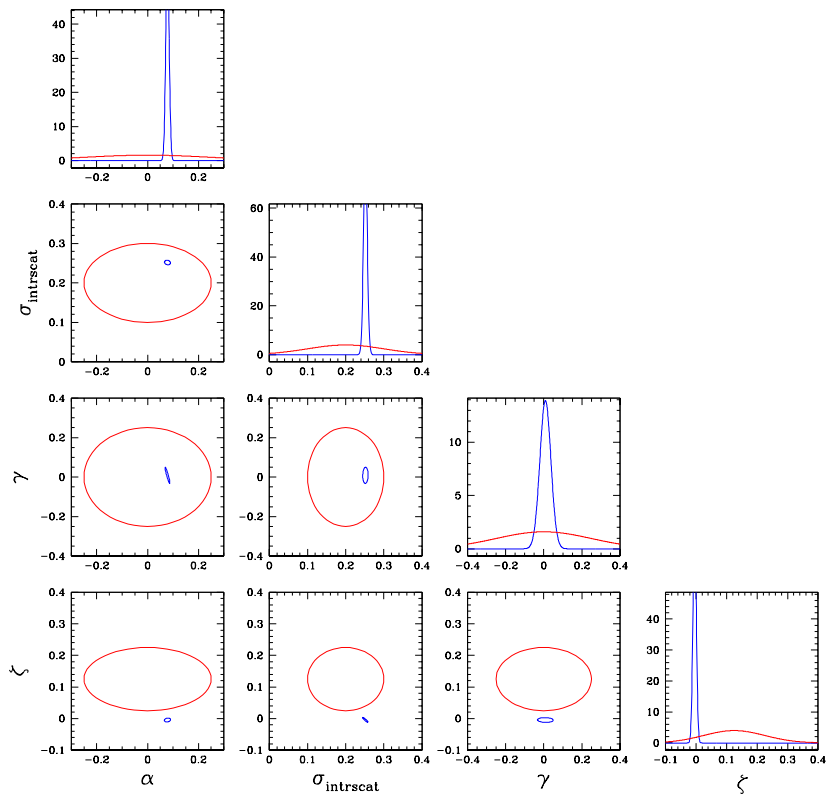,width=12truecm,clip=}
\caption[h]{Example of mismatch between predicted prior of the richness-mass 
scaling, as derived by us (in blue) and as adopted in other works (in red,
from Carbone et al. 2012). This comparison should be seen as indicative only,
because of differences between the two mass-proxy modellings. 
}
\end{figure*}

\begin{table}
\caption{Predicted richness-mass prior parameters for a PS1+Euclid-like 
survey: covariance matrix $\sigma_{i,j}$}
{
\scriptsize
\begin{tabular}{r r r r r r}
\hline
	& $\alpha$ \quad & $\beta$ & $\sigma$ & $\gamma$ & $\zeta$ \\  
$\alpha$ &$5.9 \ 10^{-5} $\\
$\beta$  &$5.4 \ 10^{-5} $& $ 2.1 \ 10^{-4}$ \\	      
$\sigma$ &$-2.1 \ 10^{-6}$& $5.9 \ 10^{-7}$ & $2.6 \ 10^{-5} $ \\    
$\gamma$ &$-2.0 \ 10^{-4} $& $-2.4 \ 10^{-4} $ & $7.5 \ 10^{-6} $ & $1.0 \ 10^{-3} $  \\	    
$\zeta$   &$1.7 \ 10^{-6} $& $-1.8 \ 10^{-7}$ & $-2.5 \ 10^{-5}$ & $-7.2 \ 10^{-6}$ & $2.9 \ 10^{-5}$   \\ 
\hline
\end{tabular}
}      
\end{table}

\section{Discussion}

Our computation of the predicted prior of the mass-richness scaling
while not accounting for
sub-dominant sources of error, such as
uncertainties related to projection and redshift-dependent errors on
the cluster photometric redshift, can easily take them
into account,
it is just a matter of replacing the assumed
likelihoods distributions (Eqs. 25 to 27) 
with the updated distributions accounting for additional
error terms one may wish to consider. For example, we can
change the normal intrinsic scatter (questioned by Shaw et al. 2010)
into a Student-$t$ distribution by typing less than ten characters (see
Appendix for details).
However, more complex simulated
data (e.g. based on an N-body simulation) are needed to generate the data
to be fitted and more and better real data are needed to characterize the real
additional dependencies (e.g. how to model the intrinsic scatter).

The starting point of literature forecasts is the end point of this paper:
they assume what our paper computes, their prior widths are our posterior
parameter uncertainties. The predicted prior is
computable and thus does not need to be assumed. Parameters show
covariance, sometimes a strong one, while none is assumed in 
literature forecasts (that we are aware of).

We note that the previous literature (starting perhaps with Lima \& Hu
2005) chose not to model the slope between mass and proxy, i.e.
implicitly assumed to know it perfectly. This assumption seems
optimistic because the slope is presently known with 25 \% accuracy (sect 2.1,
summarized in eq 19). Sect 2.3 shows that it will be known after
PS1+Euclid with a per cent accuracy. If a perfect knowledge of the slope
is assumed, then uncertainties on the other scaling parameters (scatter,
intercept, and their evolution) will be underestimated. Furthermore,
while the quality of a mass proxy is lower at the ends of the calibration
range because of the slope uncertainty, the choice performed in previous literature
works makes it a constant quality at all masses, including those outside
the range of the calibration sample. 

As mentioned above, most previous works (e.g. Lima \& Hu 2005, Cunha \& Evrard
2010, Thomas \& Contaldi 2011, Carbone et al. 2012, etc.) adopted priors for
the mass-observable scaling largely by guessing how well the relation is (or
will be) known, instead of computing the prior width.  Sometimes, the
prior width on some key parameters, like the scatter, was taken to be zero.
Some works (e.g. Cunha \& Evrard 2010, Oguri \& Takada 2011) explored the
sensitivity of cosmological constraints on the adopted priors for the
mass-observable scaling, sometimes calling this sensitivity 
``systematics", quantifying the (obvious) fact that poorly calibrated
scaling relations give poor cosmological constraints.  For example,
Cunha \& Evrard (2010) showed that cosmological constraints easily
deteriorate by a factor from $\sqrt{2}$ to $2$ by changing the prior width
from zero to $\sim 1$ \%. 

Even more important, previous forecasts did not use the information content
in the weak-lensing masses to calibrate the mass-observable scaling.
For comparison, we consider 
the priors assumed in Carbone et al. (2012), who also considered 
the mass-richness scaling of a Euclid-like survey, but made
no use of the Euclid weak-lensing masses to calibrate the mass-richness
scaling. Before proceeding
in this comparison, we emphasize a technical difference: the two
modellings are identical {\it after} swapping observable and mass variables. For
example, we modelled the scatter in proxy at a given mass as Gaussian,
while Carbone et al. (2012) modelled the scatter in mass at a given proxy as
Gaussian. Since the Carbone et al. (2012) model has no slope parameter,
for the purpose of this comparison only, we removed the slope
from the modelling (freezing it at the true value).

Fig 10 compares the prior adopted in Carbone et al. (we emphasize once more the
variable swapping) with our predicted prior.  A
major point emerges: the parameter volume encompassed by the Carbone et al.
prior, which does not use weak-lensing to calibrate the mass-richness
scaling, is $10^5$ times
larger (in the $\alpha-\sigma-\gamma-\zeta$ space) than the one we derive 
using Euclid weak-lensing masses. Similarly,
the Euclid imaging consortium science book (EICSB, Refregier et al. 2010) 
does not use the Euclid weak-lensing masses to calibrate the mass-richness
scaling and assumes a 25 \%, or 0.25, prior
uncertainty on each parameter of an observable-mass relation modelled with
a third, different, parametrization. At face value, given that our precisions
are typically one order of magnitude better per parameter, using 
weak-lensing masses may allow us to improve the knowledge of the observable-mass
scaling by a similarly large ($\approx 10^5$) amount. If the
mass-proxy scaling can be computed $10^5$ times better, 
stronger cosmological constraints can probably be inferred 
and this may alter the balance between the 
cosmological constraints achievable using cluster counts, BAOs, 
SNae, and weak-lensing tomography. Indeed, Carbone et al. (2012) estimated
that if the regression parameters
were perfectly known, then cosmological constraints tighten 
(technically: the volume of cosmological parameter 
space enclosed by the posterior probability distribution decreases)
by a factor $\approx 100$ compared to the case where one
marginalises over their (extremely wide) prior (see their Table 2). 
The gain on the constraints
on dark energy parameters alone is instead only mild: a factor 2. 
The precise computation of the gains in our specific case is, however, 
outside the aim of this work.

\section{Summary}

The aim of this work was threefold: first, using 53 clusters 
with individual measurements of
mass, we derived the richness-mass scaling in the local Universe. We
found a $0.46\pm0.12$ slope and a $0.25\pm0.03$ dex scatter in (log)
richness at a given mass measuring the richness following 
the Andreon \& Hurn (2010)
prescriptions. The fit accounts for the fact that the cluster
sample is X-ray selected and massive clusters are
over-represented, although we found that the sample selection
is a minor source of concern for this sample. Because
the scatter around the regression is derived from 
measurements of the individual masses and richnesses, our measurement
of the scatter is preferable to others derived without knowledge of
individual cluster masses, such as those of the maxBCG team (e.g.
Rykoff et al. 2011).

Secondly, using 250 $0.06<z<0.9$ clusters with spectroscopic redshift,
mostly at $z<0.3$, 
we found that the cluster redshift can be derived with an accuracy
with better than 
$\Delta z=0.02$  from the colour of the red sequence.

Thirdly, we computed the predicted prior between mass 
and richness, i.e. one of the input ingredients to judge how strongly future
surveys using clusters may constrain cosmological parameters, and to which
extent clusters can compete with other cosmological
probes.

To this aim, we generated a simulated universe obeying the derived
richness-mass scaling, observed it with a mock PanStarrs 1+Euclid-like
survey, allowing for intrinsic scatter between regressed quantities, 
allowing for mass and richness errors, and also allowing for
cluster photometric redshift
errors. The generated sample does not contain any cluster with
an $S/N>5$ weak-lensing detection at $z>0.62$  (Fig 7).

We fitted the observations with an evolving richness-mass scaling
with five parameters to be determined. We allowed an evolution in the
intercept (sometime called bias) and intrinsic scatter. We allowed an
uncertainty on the intrinsic scatter and on the intercept, as previous
works, but in contrast to all previous approaches, we did not
sidestep the modelling of the slope.  

Our fitting model recovers the input parameters, but only if the
cluster mass function and the redshift-dependent $S/N>5$ weak-lensing
survey selection function are accounted for. Neglecting them causes
fit values to deviate by $>5\sigma$ from the input  values, as a
result of the neglected Malmquist-like bias. This result
emphasizes the limitations of often adopted simplifying assumptions,
such as mass-complete redshift-independent samples.
Including the optical selection function is unnecessary because
all clusters with a weak-lensing signature are so massive and rich
that detecting their red galaxy overdensity is trivial. 
Already available imaging data from
PanStarrs 1 are of sufficient quality to detect these galaxies,
whereas mass estimates await the Euclid mission.

We derived the uncertainty and the covariance matrix of the (evolving)
richness-mass scaling, which are the input ingredients of every
cosmological forecast using cluster counts.  These five parameters will
be known with percent precision thanks to masses estimated from Euclid data. 
There are
non-negligible covariance terms between the five regression parameters.
These numbers, listed in Table 1, are the third main result of this
work.  Their determination does not require the success, or acquisition,
of other data presently not available, which is requested for other
cluster surveys, such as the XXLS and SPT survey. 

We found that the richness-mass scaling parameters can be
determined $10^5$ better (the volume enclosed by the posterior is
$10^5$ times smaller) than estimated before without 
using of weak-lensing mass estimates, although 
we emphasize that this number was derived using
scaling relations with different parametrizations. 
A better knowledge of the scaling
parameters likely has a strong impact on the relative importance of
the different probes used to constrain cosmological parameters. 

Finally, we checked that if the intrinsic scatter between
mass and richness increases by
a factor two by $z=0.6$, we are nevertheless able to recover 
the mass-richness scaling without bias, with only a factor 9 (about 1.5 per parameter) 
degradation in the quality with which
we are able to recover the scaling parameters.

The fitting code, inclusing of the treatment of the mass function
and the weak-lensing
selection function, is provided in the appendix. It can also be
re-used, for example, to derive the predicted prior of other
observable-mass scalings, such as the $L_X$-mass relation.

\begin{acknowledgements}
We thank C. Carbone, B. Sartoris, and C. Strege for their 
comments on an early version of this draft and the referee for his/her
request of clarifying our presentation concerning the impact of selection
effects in the determination of the mass-observable scaling.
\end{acknowledgements}

\appendix

\section{Model for computing the predicted prior of the
mass-richness scaling including the weak-lensing selection function}

Eq 12 and 17 to 29 are almost literally translated into JAGS (Plummer 2008),
Poisson, normal, and uniform distributions become
{\texttt{dpois, dnorm, dunif}}, respectively. 
JAGS\footnote{http://calvin.iarc.fr/$\sim$martyn/software/jags/}, following BUGS (Spiegelhalter et al. 1995), uses 
precisions, $prec = 1/\sigma^2$, in place of variances $\sigma^2$. 
The only complication comes from sampling from a distribution unavailable
in JAGS, a truncated Schechter function. This is achieved by exploiting the
property that
a Poisson($\phi$) observation of zero has a likelihood
$e^{-\phi}$. Conseguently, if our observed data are a set of 0's, and  $\phi[i]$  is
set to $- \log \mathcal{L}[i] $, we obtain the correct likelihood
contribution. The quantity $\lambda[i]$ should always be greater
than $0$, because it is a Poisson mean,
and we may accordingly need to add a suitable constant, {\texttt{C}}, 
to ensure that it is
positive. The quantity {\texttt{lg10tot.norm}} normalises the integral of the 
{\texttt{obslgM200}} likelihood to one.
The model (set of equations) reads in JAGS:

\vfill\eject \quad

{\footnotesize
\begin{verbatim}
data
{
preclgM200 <- 1./(errlgM200^2)
# normaliz 
lg10tot.norm <-0.386165-3.92996*obsz-0.247050*obsz^2-2.55814*obsz^3-5.26633*obsz^4
# dummy variable for zero-trick, to sample from a distribution not available in JAGS
for (i in 1:length(obslgM200)) {
dummy[i] <-0
}
C<-2
}
model
{
intrscat ~ dnorm(0.25,1/0.03/0.03)
prec.intrscat <- 1/intrscat^2
alpha ~ dnorm(0.08,1/0.04/0.04)
beta ~ dnorm(0.47,1/0.12/0.12)
gamma ~dt(0,1,1)
csi ~dt(0,1,1) 
for (i in 1:length(obsn200)) {
  # modelling lgM200
  # dummy prior, requested by JAGS, to be modified later
  lgM200[i] ~ dunif(13.9891+1.04936*obsz[i]+0.488881*obsz[i]^2,16)
  # modelling a truncated schechter
  lnnumerator[i] <- -(10^(0.4*(lgM200[i]-12.6+(obsz[i]-0.3))))
  # its integral, from the starting point of the integration (S/N=5)
  loglike[i] <- -lnnumerator[i]+lg10tot.norm[i]*log(10)+C   
  # sampling from an unavailable distribution
  dummy[i] ~ dpois(loglike[i])
  obslgM200[i] ~ dnorm(lgM200[i],preclgM200[i])
  # modelling n200, z and relations
  obsn200[i] ~ dpois(pow(10, lgn200[i])) 
  obsz[i] ~ dnorm(z[i],pow(0.02,-2)) 
  z[i]~dnorm(0,1)
  # modelling mass -n200 relation allowing for evolution
  lgn200m[i] <- alpha+1.5 +beta*(lgM200[i]-14.5)+ gamma*(log(1+z[i]))
  lgn200[i] ~ dnorm(lgn200m[i], prec.intrscat.z[i])	  
  prec.intrscat.z[i] <- 1/( 1/prec.intrscat-1+(1+z[i])^(2*csi))
  }
}
\end{verbatim}
}

To adopt a Student's $t$--distribution with ten degrees of freedom 
{\texttt{dt}} to model the intrinsic scatter (Sect. 4),
it suffices to replace the line starting by {\texttt{lgn200[i]}} 
with

 {\footnotesize
\begin{verbatim}
 lgn200[i] ~ dt(lgn200m[i], prec.intrscat.z[i],10)
\end{verbatim}
}

\end{document}